\documentclass{article}
\usepackage{spconf,amsmath,graphicx,hyperref}
\AtBeginDocument{%
  }
\usepackage[table]{xcolor}
\usepackage{amsmath}  % 
\usepackage{booktabs} 
\usepackage{array}    % 支持 m 列类型，垂直居中
\usepackage{multirow}
\usepackage{graphicx}
\usepackage{amsfonts}

\title{UTI-LLM: A Personalized Articulatory-Speech Therapy Assistance System Based on  Multimodal Large Language Model}
\name{Yudong Yang$^{1}$, Xiaokang Liu$^{1}$, Shaofeng zhao$^{3}$, Rongfeng Su$^{1}$, Nan Yan$^{1,2,*}$, Lan Wang$^{1,2,*}$\thanks{*Corresponding authors}}
\address{$^{1}$Shenzhen Institutes of Advanced Technology, Chinese Academy of Sciences, China\\
$^{2}$Key Laboratory of Biomedical Imaging Science and System, Chinese Academy of Sciences, China \\
$^{3}$Department of Rehabilitation Medicine,\\The Eighth Affiliated Hospital of Sun Yat-sen University, China\\
}

\begin{document}
\ninept
\maketitle
\begin{abstract}
Speech therapy is essential for rehabilitating speech disorders caused by neurological impairments such as stroke. However, traditional manual and computer-assisted systems are limited in real-time accessibility and articulatory motion feedback. Recent advances in multimodal large language models (MLLMs) have demonstrated significant potential in healthcare, especially through their adaptive assessment and therapeutic feedback capabilities. Nevertheless, challenges including insufficient acquisition and fusion of articulatory information, inadequate parsing of articulatory organ motion trajectories, and the scarcity of domain-specific datasets hinder the application of MLLMs in speech therapy. To address these limitations, we propose an MLLM-based speech rehabilitation assistance system that  leverages ultrasound tongue imaging and speech signals to deliver precise, interactive articulatory feedback. We construct a high-quality domain-specific dataset comprising ultrasound-speech dialogue pairs. This dataset facilitates fine-tuning to enhance the model's clinical adaptability. Furthermore, our method develops spatiotemporal fusion training strategy of ultrasound videos and speech signals, enabling fine-grained articulatory impairment analysis and ultimately generating actionable feedback. Experimental results demonstrate the effectiveness of our model in articulatory analysis and clinical assessment. 
\end{abstract}
\section{Introduction}
\label{sec:intro}
Speech disorders, such as dysarthria caused by conditions like stroke, represent debilitating sequelae that severely impact patients' quality of life \cite{latif2020speech,freed2023motor}.  Clinical evidence demonstrates that without timely intervention, these conditions often progress to chronic states, markedly complicating subsequent rehabilitation. However, currently speech therapies face dual challenges: a critical shortage of qualified therapists due to the extended training periods and high costs of professional development \cite{marante_school-based_2023}, which hinders accessible care for many patients \cite{squires_addressing_2013}. These challenges underscore the urgent need for computer-assisted systems capable of objective assessment and speech rehabilitation.

While preliminary explorations into objective assessment and rehabilitation methods for speech disorders exist \cite{usha2023speech, strand2013motor, liu2025automatic, guo2025mbbo}, their clinical translation remains hindered by persistent bottlenecks. The core challenge lies in the integration of multimodal pathological data to enable personalized therapeutic strategies and real-time feedback. Although speech modalities remain central to speech rehabilitation, medical data also provide complementary insights. Furthermore, Ultrasound tongue imaging (UTI), as a non-invasive, portable technology enabling real-time sagittal visualization of tongue dynamics during continuous speech, has emerged as a critical tool for speech disorder assessment \cite{ribeiro2021exploiting, preston2017ultrasound, yang2024feature666}. Regrettably, current rehabilitation systems predominantly rely on uni-modal analysis, failing to exploit the potential of cross-modal information.

The clinical utility of rehabilitation systems hinges on two key factors: interactive adaptability and linguistic naturalness. An ideal system must balance clinical with fluid human-machine interaction. Existing solutions, such as Computer-Aided Speech and Language Therapy \cite{oster2002presentation, popovici2012professional, das2017automated}, predominantly employ retrieval-based dialogue models that struggle to achieve authentic conversational flow. These systems exhibit limitations: (1) restricted multimodal fusion capabilities, often focusing on unimodal (e.g., audio ) while neglecting complex articulatory-acoustic-kinematic interactions, and (2) inadequate assessment of speech disorder manifestations and lack of conversational explanatory functions, failing to meet clinical demands for personalized feedback.

Recent advancements in methodologies leveraging multi-modal have revolutionized large language model (LLM) development\cite{li2023llava, guo2024llava, meng2025large}. In healthcare, MLLMs have demonstrated significant potential across medical imaging analysis and clinical decision support \cite{el2024democratizing, alsaad2024multimodal}. Nevertheless, their cross-modal semantic alignment capabilities remain insufficient for speech rehabilitation demands. Current MLLMs primarily trained on open-domain web data \cite{tang2024avicuna, maaz2023video} exhibit limited comprehension when processing specialized medical modalities like UTI and articulatory tongue motion patterns. Specifically, three critical challenges persist: 1) \textbf{Limited Ultrasound Data Comprehension}: Existing models may not interpret tongue motion patterns and articulatory features in ultrasound scans. 2) \textbf{Difficulty in Temporal Information Fusion:} There is a lack of effective mechanisms to integrate high-dimensional ultrasound videos with time-series audio and kinematic signals, and 3)\textbf{ Scarcity of Multimodal Data: }The absence of large-scale and annotated medical multimodal Question-Answer (QA) datasets restricts model generalization and transferability in speech rehabilitation tasks.

To address these challenges, we propose a MLLM-based speech rehabilitation dialogue system with three key contributions:  \textbf{(1).Multimodal Instruction Tunging Dataset}: We developed an include 10k high-quality sample multimodal dataset for speech rehabilitation understanding dialogue, constructed from annotated UTI–speech parallel data. \textbf{(2).UTI-LLM Architecture}: We propose a MLLM capable of processing both UTI and speech signals as input, to deliver personalized therapeutic recommendations through interpretable reasoning of tongue movement and providing dysarthria assessment and analysis.  \textbf{(3).Temporal-Spatial Dynamic Understanding UTI-Speech}: We are the first to use UTI–speech parallel data for speech rehabilitation reasoning dialogues. we utilize fusion speech signals, UTI spatial and temporal motion features, enhancing the model attention to tongue movements in UTI.

\section{Methods}

\subsection{Data Preparation and QA Dialogue Construction}
\textbf{Tongue Movement Trajetcory Preparation for UTI}: Tongue movement trajectories play a critical role in understanding speech personalized rehabilitation processes. To enable the LLM to learn and accurately and explainability describe tongue movement information, we first construct a annotating dataset of tongue movement trajectories. An ultrasound sample is defined as $x_t$ and its corresponding database $D_{UTI} = \{x_1, \ldots, x_T\}$. To extract representative frames, we apply a clustering algorithm that groups frames with high similarity. We have $K$ clusters $C_1, C_2, ..., C_K$  and each cluster $C_n$ contains $n_k$ data points, with the cluster center denoted as ${\mu}_k$.

Then, for the initial manual annotation, three key regions of the tongue are defined: tip, body and root of the tongue, denoted $R_1^P,R_2^P,R_3^P$, where $P$ represents the number of points in each region. Each ultrasound frame is annotated with $P$ key trajectory points, ensuring that their spatial coordinates satisfy the constraint, which represent the trajectory variations of different tongue regions: 
\begin{equation}
\forall i<j,x\left(R_i^P\right)<x\left(R_j^P\right)
\end{equation}

\begin{figure}
\includegraphics[width=0.5\textwidth]{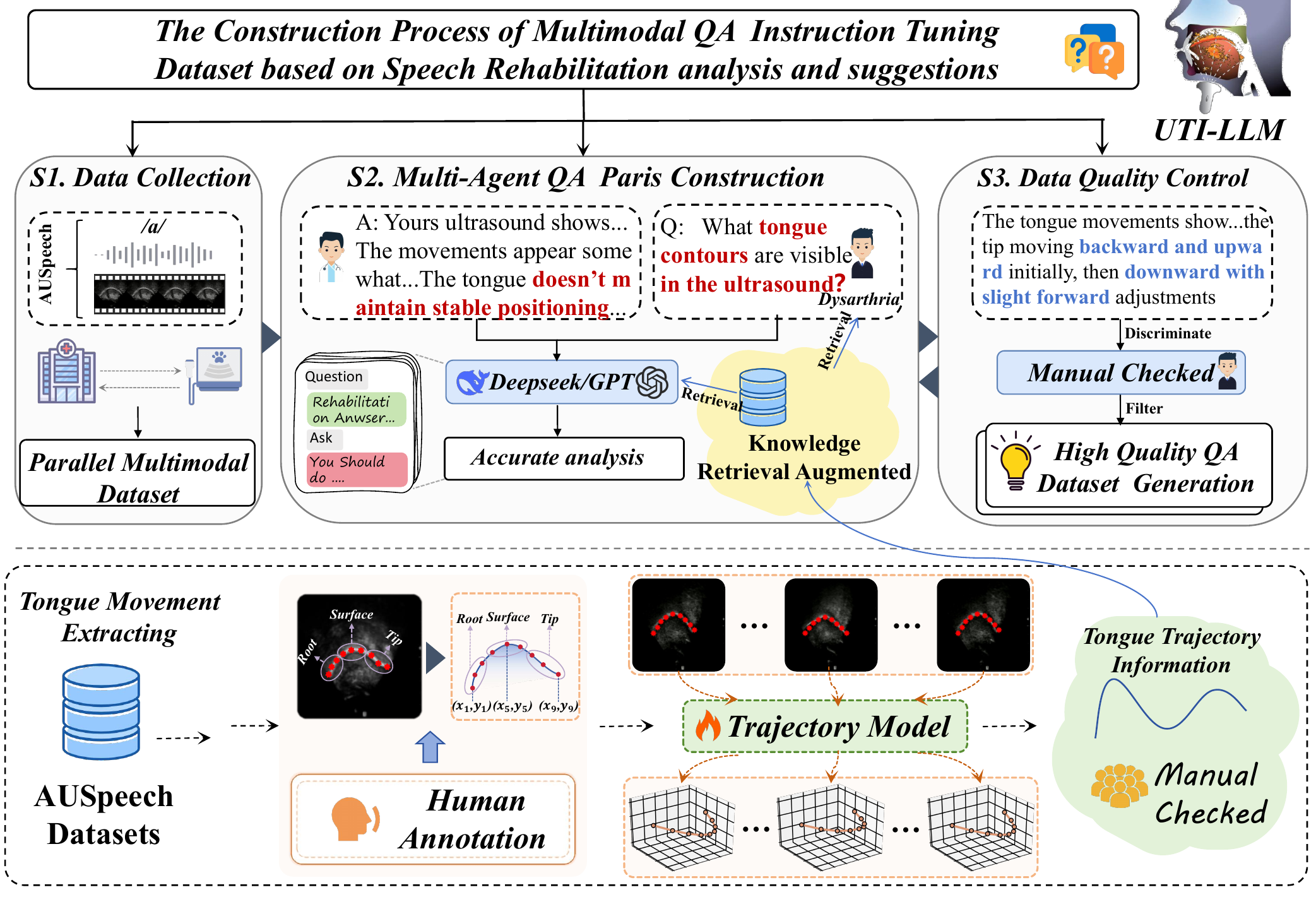}
\caption{Agent QA Framework for Medical Dialogue Generation. The schematic illustrates the interactive pipeline of our datasets, which simulates real-world clinical consultations.}
\label{fig:fig1}
\end{figure}

Additionally, we have developed a Resnet-Based \cite{he2016deep} trajectory extraction model designed to automatically identify key tongue movement using labeled datasets. This approach reduces the cost of manual annotation and ultimately yields a dataset verified by experts to ensure quality. Given an input UTI frame \(\mathbf{x}_i\), we first extract its visual features through a feature extraction module $\mathcal{F}$: 
\begin{equation}
    \mathbf{F}_i = \mathcal{F}(\mathbf{x}_i) \in \mathbb{R}^{h \times w \times c}, 
\end{equation}
where \(\mathbf{F}_i\) denotes the extracted feature map with spatial dimensions \(h \times w\) and \(c\) channels. and the tongue region coordinates \(\widehat{\mathbf{P}}_i\) are predicted via a regression head \(\mathcal{G}\): $\widehat{\mathbf{P}}_i = \mathcal{G}(\mathbf{F}_i)$.\\
\textbf{Dual-Agent Collaborative QA Generation Framework:} To address the need for generating dialogue data, this paper proposes an agent strategy for constructing high-quality medical QA pairs through a modular design. As illustrated in Figure~\ref{fig:fig1}, the framework includes a Doctor Agent and a User Agent. These agents iteratively refine responses through a dynamic knowledge retrieval mechanism: (1) These agent generates initial prompts based on predefined task templates, covering dysarthria assessment, tongue motion analysis, and rehabilitation advice.(2) The agent is equipped with a knowledge inference capabilities, enabling real-time access to ultrasound trajectory data, pronunciation databases, and diagnostic information. This ensures that generated responses maintain medical accuracy.
\\
\textbf{User Agent: Question Generation Strategy: }The User Agent is responsible for proactively generating questions based on predefined topic types. To enhance data diversity, we introduce a temperature-based sampling strategy to control the randomness of generated questions. Let $\tau \in [0.1, 1.0]$ denote the temperature parameter.

To avoid semantic convergence caused by data pattern homogeneity, we define a diversity constraint function $\mathcal{D}(Q)$ that quantifies the semantic novelty of a generated question $Q$ with respect to the existing dataset $\mathcal{H}$:
\begin{equation}
    \mathcal{D}(Q) = 1 - \max_{q_i \in \mathcal{H}} \text{Sim}(Q, q_i)
\end{equation}

\vspace{-5.5px}
Here, $\text{Sim}(\cdot)$ denotes a semantic similarity function. If $\mathcal{D}(Q)$ falls below a predefined threshold, the system will resample or reconstruct the question.\\
\textbf{Doctor Agent: Response Generation Mechanism}
Upon receiving a question from the User Agent, the Doctor Agent retrieves structured medical knowledge to generate informed and medically grounded responses. The knowledge base contains four key components:(1)Tongue trajectory coordinates: $\mathcal{T} = \left\{(x_t, y_t)\right\}_{t=1}^{T}$; (2)Phonetic type: $\mathcal{P}_{\text{type}}$ (e.g., Monophthong); (3)Phonetic content: $\mathcal{P}_{\text{text}}$ (e.g., /a/ ); (4) Diagnostic label: $\mathcal{D}_{\text{diag}}$ (speech disorder label)

Given the central role of trajectory comprehension, we apply normalization to the raw trajectory sequence $\mathcal{T}$ obtained from the tracking model, mapping it into a unit space:

\begin{equation}
    \widetilde{\mathcal{T}} = \left\{ \left( \frac{x_t - x_{\min}}{x_{\max} - x_{\min}}, \frac{y_t - y_{\min}}{y_{\max} - y_{\min}} \right) \right\}_{t=1}^{T}
\end{equation}

This transformation ensures that the model focuses on the pattern of motion rather than absolute coordinates influenced by individual anatomical differences.Furthermore, we introduce a motion amplitude threshold $\delta$ and retain only those trajectory samples that meet the following condition: $\max_{t \in [1, T]} \left| (x_t, y_t) - (x_1, y_1) \right|_2 > \delta$

Furthermore, to ensure responses are logical and knowledge-grounded, the model must condition its generation on retrieved information. The final output is produced through a multimodal reasoning function that integrates relevant knowledge:

\begin{equation}
    R = f_{\text{LLM}}\left( \widetilde{\mathcal{T}} \oplus \mathcal{P}_{\text{type}} \oplus \mathcal{P}_{\text{text}} \oplus \mathcal{D}_{\text{diag}} \right)
\end{equation}

Here, $\oplus$ denotes the concatenation or fusion of multimodal inputs, and $f_{\text{LLM}}(\cdot)$ represents LLM inference to jointly reason over trajectory, pronunciation information, and diagnostic result.\\

\begin{figure}
\includegraphics[width=0.5\textwidth]{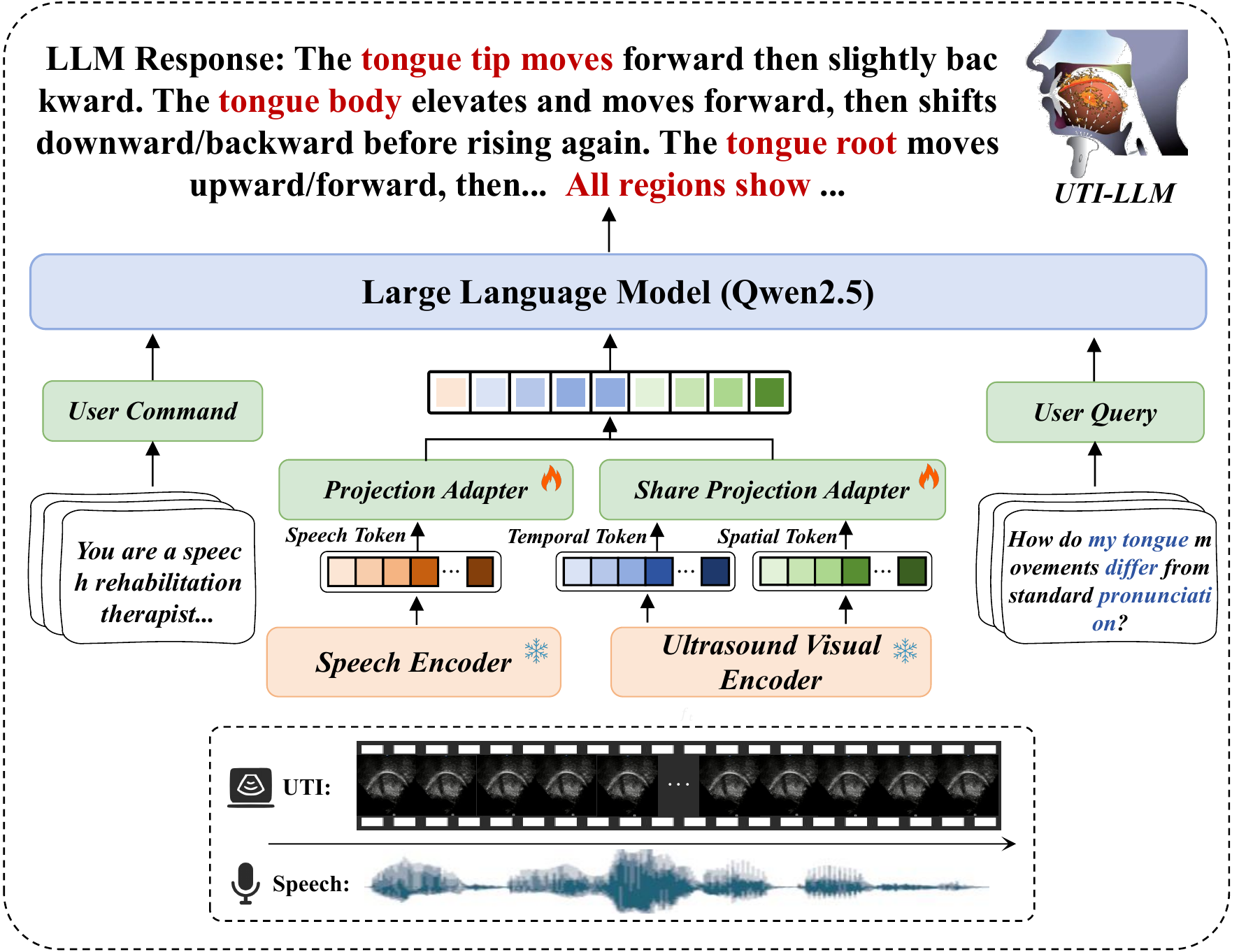}
\caption{Architecture of UTI-LLM Multimodal Speech Rehabilitation. The architecture enables joint reasoning over articulatory movement, acoustic and UTI visual cues for speech rehabilitation.}
\label{fig:figures2}
\end{figure}
\vspace{-3px}
\subsection{Mainstream Architecture}
Existing approaches often process ultrasound and speech modalities in isolation, failing to capture critical articulatory-acoustic interactions essential for speech rehabilitation. To effectively integrate these complementary modalities for comprehensive pathological analysis, we need dynamic alignment between articulatory movements and acoustic patterns, capturing pathological correlations essential for rehabilitation assessment. The specific architecture of our method is as follows:\\
\textbf{Instruction Tuning}:
As illustrated in Figure~\ref{fig:figures2}, We use the Qwen2.5-7B~\cite{chu2024qwen2} as our LLM to process audio-visual tokens and user queries, LLM generating responses $\mathbf{R}$:
\begin{equation}
    \mathbf{R} = LLM(<Instruction> <Q_{UTI}> <Q_{Speech}>)
\end{equation}
Here, $<Instruction>$ is a task-specific prompt selected from a medical QA template pool, for example articulation correction, tongue motion feedback etc. $<Q_{UTI}>$ and $<Q_{Speech}>$ are the encoded ultrasound and speech features, respectively. The model is trained to maximize the likelihood of generating the appropriate $R$.\\
\textbf{Speech Encoder and Connector Adapter:}  
We employ hubert \cite{hsu2021hubert} to extract speech features, prioritizing the final layer (layer $L$) . This choice reduces acoustic variability (e.g., background noise) while preserving speaker-specific attributes. Given raw speech input $\mathbf{A} \in \mathbb{R}^{T_a}$, where $T_a$ is the sequence length of raw audio samples (with subscript $a$ denoting audio-related terms), the hubert encoder produces frame-level hidden states $\mathbf{H} = [\mathbf{h}_1, ..., \mathbf{h}_n] \in \mathbb{R}^{n \times d_h}$ for $n$ frames. We select the penultimate layer outputs $\mathbf{H}^{L-1}$ and apply a learnable projection $\mathbf{W}_a \in \mathbb{R}^{d_h \times d_e}$ to align speech features with the language decoder's embedding space: $\mathbf{Q}_a = \mathbf{H}^{L-1} \cdot \mathbf{W}_a \in \mathbb{R}^{n \times d_e}$.\\
\textbf{UTI Visual Encoder and Connector Adapter:}  
For video inputs $\mathbf{V} \in \mathbb{R}^{T_v \times H \times W \times C}$ with $T_v$ frames (with subscript $v$ denoting videl-related terms), resolution $H \times W$, and $C$ channels, we use a pretrained CLIP ViT-L/14 model \cite {radford2021learning}. Each frame is split into $N = (H/p) \times (W/p)$ patches ($p=14$), generating frame embeddings $\mathbf{X} \in \mathbb{R}^{T_v \times N \times d}$. 

Here, we employ temporal and spatial features to comprehensively capture both the spatial configuration and dynamic motion patterns of the tongue in UTI. This design enables the model to perceive not only spatial articulatory postures but also the temporal evolution of tongue movements The specific operations are as follows:
(1) Spatial Understanding Token: Average embeddings across time steps for each spatial patch:
\begin{equation}
    \mathbf{z}_i = \frac{1}{T_v} \sum_{j=1}^{T_v} \mathbf{X}[j,i] \in \mathbb{R}^{N \times d}.
\end{equation}
\vspace{-3px}
(2) Temporal Trajectory Token: Average embeddings across spatial patches for each frame: 
\begin{equation}
    \mathbf{t}_i = \frac{1}{N} \sum_{j=1}^N \mathbf{X}[i,j] \in \mathbb{R}^{T_v \times d}
\end{equation}
\vspace{-3px}
The concatenated features $\mathbf{v} = [\mathbf{t}; \mathbf{z}] \in \mathbb{R}^{(T_v+N) \times d}$ are projected via a linear layer $\mathbf{W}_v \in \mathbb{R}^{d \times d_e}$: $
    \mathbf{Q}_v = \mathbf{v} \cdot \mathbf{W}_v \in \mathbb{R}^{(T_v+N) \times d_e}.
$

\section{Experiments}
\subsection{Implementation Details}
We trained a 7B-parameter model on four NVIDIA A6000 GPUs. The configuration included 50 epochs, a learning rate of $1\times10^{-5}$, and Low-Rank Adaptation(Lora) with rank \texttt{r} = 64 and \texttt{alpha} = 128. We applied a $1\times10^{-6}$ warmup lr and used weight decay $0.05$. The maximum sequence length was set to 1024 tokens. To remove low-motion regions, we performed K-means clustering on extracted keyframes, keeping 100 clusters and discarding frames with low dynamic variance. For multi-agent dataset construction, we utilized both DeepSeek-V3-671B \cite{liu2024deepseek} for dialogue generation.

\vspace{-2px}
\subsection{Datasets}
We utilize UTI-speech parallel data from the AUSpeech dataset~\cite{auspeech}, the dataset includes 43 normal and 11 patient speakers with dysarthria and total duration of 22.31 hours. we focus specifically on healthy speaker session1 and patients speaker session. The dataset spans four articulation tasks: monophthong, monosyllables, Sentence, and swallowing, which provides a comprehensive representation of tongue motion dynamics and speech variability, and no overlap exists between the train and test sets.

\vspace{-2px}
\subsection{Evaluation Metrics}
We evaluate from three aspects:(1) \textbf{UTI Tongue Movement Analysis}: We assess the correctness of the LLM-generated tongue movement description for UTI using standard natural language generate(NLG) metrics, including BLEU, METEOR, and ROUGE-L.(2) \textbf{Dysarthria Assessment}: The LLM's ability to classify dysarthria is evaluated using accuracy annd F1-score.(3) \textbf{Human Experts and LLM-based Evaluation}: LLM is used to score model responses along three dimensions:  (i) Correctness, which measures the factual accuracy of the output content; (ii) Trajectory Consistency, which evaluates the consistency of the tongue trajectory descriptions in the output; and (iii) Completeness, which assesses whether the output content is comprehensive. Additionally, 3 human linguistics experts blind evaluate the outputs based on content consistency, generate correctness, and practical usefulness.

\begin{table*}[ht]
\centering
\caption{Comparison of Different Methods on natural language generate metrics average Scores.}
\label{tab:table2}
\resizebox{1.0\linewidth}{!}{%
\begin{tabular}{lccccccccc}  
\toprule
Method & Visual  & Audio & BLEU-1$\uparrow$ & BLEU-2$\uparrow$ & BLEU-3$\uparrow$ & METEOR$\uparrow$ & ROUGE-L $\uparrow$  & AVERAGE SCORES$\uparrow$ \\
\midrule
Video-Chatgpt \cite{maaz2023video} & \checkmark & \texttimes   &0.3778 & 0.2188 & 0.1322 & 0.3277 & 0.3641 & 0.2841 \\
Qwen2-Audio \cite{chu2024qwen2} & \texttimes & \checkmark & 0.4649 & 0.3115 & 0.2336 & 0.4171 & 0.4215 & 0.3697\\
PandaGPT \cite{su2023pandagpt} & \checkmark & \checkmark  & 0.4749 & 0.3310 & 0.2460 & 0.4422 & 0.4157 & 0.3820\\
Avicuna \cite{tang2024avicuna} & \checkmark & \checkmark  & 0.4165 & 0.2895 & 0.2126 & 0.4062 & 0.3774 & 0.3404\\
\midrule
UTI-LLM (ours) & \checkmark & \checkmark  & \textbf{0.4845} & \textbf{0.3442} & \textbf{0.2654} & \textbf{0.4660} & \textbf{0.4367} & \textbf{0.3994}\\
\bottomrule
\end{tabular}
}
\end{table*}

\begin{table}[ht]
\centering
\vspace{-3mm}
\caption{Evaluation of model Assessment ability for dysarthria}
\label{tab:table1}
\resizebox{0.85\linewidth}{!}{%
\begin{tabular}{lcccccccc}  
\toprule
Method & Accuracy$\uparrow$ & F1-Scores$\uparrow$ & Average$\uparrow$ \\
\midrule  
Video-Chatgpt  & 0.6855 & 0.6173 & 0.6514 \\
PandaGPT  & 0.6546 & 0.6735 & 0.6641  \\
Avicuna  & 0.7139 & 0.7329 & 0.7234\\
Qwen2-Audio & 0.7835 & 0.7145 & 0.7490   \\
Ours  & \textbf{0.9098} & \textbf{0.9058} & \textbf{0.9078} \\
\bottomrule
\end{tabular}%
}
\end{table}

\begin{table}[ht]
\centering
\caption{LLM-Based Evaluation of Model Output Understanding}
\resizebox{1.0\linewidth}{!}{%
\begin{tabular}{lcccc}  
\toprule
Method & Correctness$\uparrow$ & Consistency$\uparrow$ & Completeness$\uparrow$ & Average$\uparrow$\\
\midrule  
Video-Chatgpt & 1.97 &  2.06 & 2.67 & 2.23 \\
PandaGPT & 2.21 & 2.15 & 2.74 & 2.37\\
Avicuna & 1.61 & 1.69 & 2.02 & 1.77 \\
Qwen2-Audio & 2.23 & 2.09 & 2.64 & 2.32 \\
Ours &  \textbf{2.76} & \textbf{2.57} & \textbf{3.22} & \textbf{2.85} \\
\bottomrule
\end{tabular}%
}
\label{tab:table1}
\end{table}

\begin{table}[ht]
\centering
\vspace{-3mm}
\caption{Linguistic human experts evaluation for difference methods}
\resizebox{1.0\linewidth}{!}{%
\begin{tabular}{lcccc}  
\toprule
Method &  Consistency $\uparrow$ & Correctness $\uparrow$ & Usefulness$\uparrow$ & Average$\uparrow$\\
\midrule  
Video-Chatgpt &  2.83 & 2.50 & 2.66 & 2.66\\
PandaGPT & 3.83 & 3.83 & 3.66 & 3.77 \\
Avicuna & 2.50  & 3.16  & 2.66 & 2.77 \\
Qwen2-Audio & 3.33  &3.16  & 3.33 & 3.27 \\
Ours & \textbf{4.00} & \textbf{4.16} & \textbf{4.00} & \textbf{4.05}\\
\bottomrule
\end{tabular}%
}
\label{tab:table1}
\end{table}

\vspace{-3px}
\subsection{Experimental Result}
Given the lack of existing MLLM designed for speech rehabilitation, we adapt different MLLM to our task by fine-tuning them on our UTI-Speech dataset, Our experimental results are as follows:\\
\textbf{Result on Tongue Movement Interpretable Reasoning } The evaluation using NLG metrics demonstrates our model superior's capability in translating complex tongue movement kinematics from ultrasound data into coherent textual descriptions. As shown in Table 1, our method outperforms the best baseline by 4.5\% in overall score. The higher BLEU and METEOR specifically indicate improved semantic alignment with expert descriptions, while the ROUGE-L score reflects enhanced syntactic completeness in describing tongue movement. This demonstrated our model capacity for motion characterization and linguistically fluent response generation.

In the LLM-Based evaluation (Table 3), our model achieves a 24\% improvement in Correctness, a 20\% gain in Trajectory Consistency, and a 22\% increase in Completeness. These results are further corroborated by human expert evaluations (Table 4), where our method attains the highest scores across all subjective metrics, including Consistency, Correctness, and Usefulness, significantly outperforming baselines. These improvements collectively show that: 1) the model exhibits stronger observational capabilities for capturing subtle kinematic features; 2) it provides more transparent and interpretable reasoning for rehabilitation suggestions; and 3) it delivers more clinically reliable and practically useful outputs for tongue movement analysis.\\
\textbf{Result on Dysarthria Assessment:} We evaluated whether our model can effectively assess dysarthria. The results demonstrate that our approach achieved a significant improvement of 16.11\% in accuracy and 26.77\% in F1-score over the Qwen2-Audio. This notable enhancement confirms our model superior capability in capturing pathological patterns in speech, indicating its strong potential for clinical applications such as automated dysarthria screening and objective assessment of speech motor disorders.\\
\textbf{Module Ablation Studies:} Our experimental results demonstrate the complementary nature of speech and UTI modalities for this task. As shown in Figure 3, the UTI-only model significantly outperforms the speech-only baseline, indicating the rich articulatory information captured by ultrasound tongue imaging. Furthermore, the multimodal Speech+UTI configuration achieves the best performance, representing a +6.0\% relative improvement over speech-only and +1.7\% over UTI-only approaches. This confirms our hypothesis that the complementary information from both modalities enables more robust feature representation, with speech providing acoustic cues while UTI contributes visualization of articulatory movements.

\begin{figure}
\centering
\includegraphics[width=0.43\textwidth]{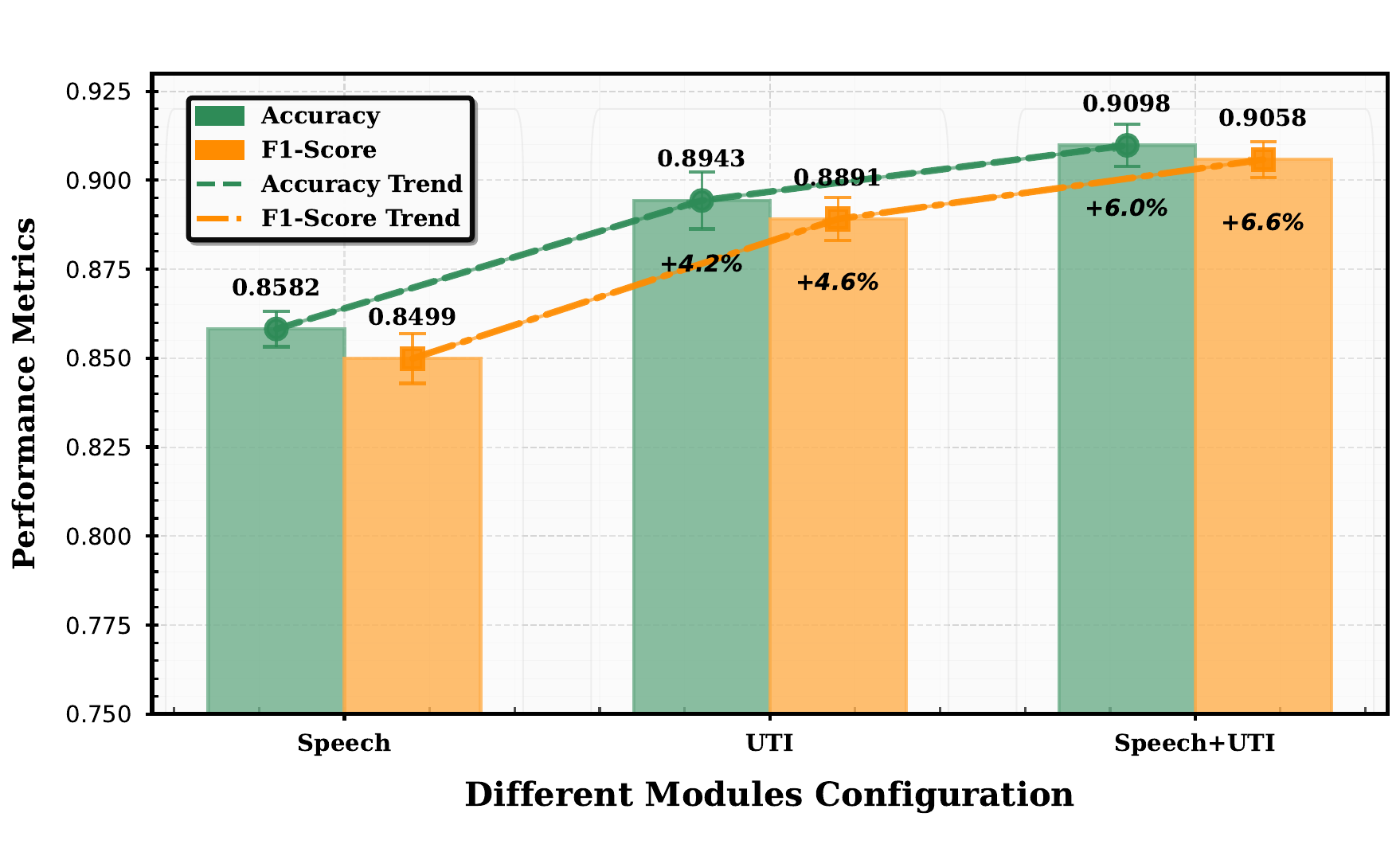}
\caption{Ablation experiments with different module configurations.(demonstrated the necessity of our UTI and speech)}
\label{fig:figures3}
\end{figure}

\section{Conclusion}
In this work, we introduced a speech rehabilitation reasoning dialogue framework that leverages the integration of UTI and speech signals. By employing MLLM, we enable cross-modal alignment, facilitating personalized therapeutic recommendations. Our curated dataset of high-quality samples, derived from UTI–speech parallel data, serves as a robust foundation for domain-specific applications. Our approach enhances the model's attention to tongue movements, significantly improving speech rehabilitation outcomes.

\section{Acknowledgements}
This research is supported by National Natural Science Foundation of China (U23B2018), National Natural Science Foundation of China (NSFC 62271477), Shenzhen Science and Technology Program (JCYJ20220818101411025, JCYJ20220818102800001, JCYJ20220818101217037), Shenzhen Peacock Team Project (KQT\\D20200820113106007), Shenzhen Medical Research Fund (C24010\\01), Shenzhen University-Huaqiang Project.

\label{sec:refs}

\bibliographystyle{IEEEbib}
\bibliography{strings,refs}

\end{document}